\documentstyle[seceq,twocolumn]{jpsj}
\input epsf
\title{Phase Diagram of $S=1$ Bond-Alternating XXZ Chains \\
Reflecting the Hidden $Z_{2}\times Z_{2}$ Symmetry}
\author{Atsuhiro {\sc Kitazawa}$^{1,2}$ 
and Kiyohide {\sc Nomura}$^{2}$}
\inst{${^{1}}$Department of Physics, 
Kyushu University, Fukuoka 812-81, Japan \\
${^{2}}$Department of Physics, Tokyo Institute of Technology, 
Oh-okayama, Meguro-ku, Tokyo 152, Japan}
\recdate{\today}
\abst{The phase diagram of 
the spin-1 bond-alternating XXZ chain is studied numerically. 
This model is appropriate to study the VBS picture 
and the hidden $Z_{2}\times Z_{2}$ symmetry concerning to 
the Haldane gap problem. 
The possible phases are the ferromagnetic, the $XY$, the Haldane, the dimer, 
and the N\'eel ones. 
The critical properties of this model can be interpreted as the
Ashkin-Teller type reflecting the hidden $Z_2\times Z_2$ symmetry. 
The phase transitions are of the Berezinskii-Kosterlitz-Thouless type 
(XY-Haldane and XY-dimer), 
of the 2D Gaussian type (Haldane-dimer), 
and of the 2D Ising type (Haldane-N\'eel and dimer-N\'eel). 
We determine the phase boundary and universality class 
using the recent developed techniques by the authors.}

\kword{Haldane gap, $S=1$ bond-alternating XXZ chain, 
Hidden $Z_{2}\times Z_{2}$ symmetry, BKT transition, Gaussian transition}

\def\runtitle{Phase Diagram of $S=1$ Bond-Alternating XXZ Chains 
Reflecting the Hidden $Z_{2}\times Z_{2}$ Symmetry}
\def\runauthor{Atsuhiro {\sc Kitazawa} and Kiyohide {\sc Nomura}}
\def\sf{\rm}

\begin{document}
\sloppy
\maketitle
\section{Introduction}
Haldane \cite{Haldane} made a fascinating conjecture 
that the integer spin $S$ Heisenberg antiferromagnetic chain 
has a unique disordered ground state with an energy gap, 
while for a half odd integer $S$
it is critical without energy gap and belongs to the same
universality class as the $S=1/2$ case.  
For $S=1$, this conjecture is practically established 
by many numerical \cite{Nightingale,White} 
and experimental \cite{Buyers,Renard,Katsumata} studies.  
Affleck {\it et. al.}\cite{AKLT} proposed the valence bond solid (VBS) 
state for the Haldane gap systems. 
They studied a $S=1$ isotropic spin chain 
with special biquadratic interactions and constructed the exact 
ground state whose spin correlation decay exponetially. 
They also showed the existance of a finite gap for the excitation. 

For this problem, den Nijs and Rommelse\cite{Nijs} proposed 
the string order parameter,
\begin{equation}
  {\cal{O}}^{\alpha}_{\mbox{\small string}} = -\lim_{|j-k|\rightarrow\infty}
  \left\langle S^{\alpha}_{j}
\exp\left[ i\pi\sum_{l=j+1}^{k-1} S^{\alpha}_{l}\right] S^{\alpha}_{k}
  \right\rangle 
\end{equation}
(where $\alpha= z,x$ and 
$\langle\cdot\rangle$ means an expectation in the ground state) 
which measures the hidden antiferromagnetic order and 
characterizes the $S=1$ Haldane phase. 
By the Goldstone's theorem, 
if some symmetry is broken in the Haldane gap system, it is not expected 
a continuous one. 
Later, the string order parameter was related to the hidden discrete
$Z_{2}\times Z_{2}$ symmetry by Kennedy and Tasaki\cite{KT3}. 
Using a nonlocal unitary transformation, they transformed the $S=1$ spin chains 
to one model which has local interactions and the discrete symmetry explicitly. 
The string order parameters are transformed to the ferromagnetic ones, 
and they relate to the hidden $Z_{2}\times Z_{2}$ symmetry. 

In our previous paper\cite{KNO} with Okamoto, we have studied the $S=1$ 
$XXZ$ spin chain with bond alternation,
\begin{equation}
  H = \sum_{j=1}^{N}(1-\delta(-1)^{j})[S^{x}_{j}S^{x}_{j+1}
  +S^{y}_{j}S^{y}_{j+1}+\Delta S^{z}_{j}S^{z}_{j+1}],
\label{ham}
\end{equation}
and found that the phase diagram reflects 
the hidden $Z_{2}\times Z_{2}$ symmetry. 
The possible phases are the ferromagnetic, the $XY$, the Haldane, 
the dimer, and the N\'eel ones (Fig.1). 
For $\delta=0$, many numerical studies have been done. 
The Haldane-N\'eel transition occurs at $\Delta_{c1}=1.17\pm0.02$ and belongs 
to the 2D Ising universality class\cite{BJ,ST,YT,KN}. 
Near $\Delta=\Delta_{c1}$, the energy gap behaves as $|\Delta-\Delta_{c1}|$, 
so the critical exponent $\nu$ is $1$. 
At $\Delta=0$ precisely, the Haldane-$XY$ transition occurs 
and it is of the Berezinskii-Kosterlitz-Thouless (BKT) type\cite{KNO}. 
For this transition, the energy gap closes as 
$\exp(-\mbox{const}/\sqrt{\Delta})$, 
and in the $XY$ phase the excitation is gapless. 
In the Haldane phase ($0<\Delta <\Delta_{c1}$) 
the hidden $Z_{2}\times Z_{2}$ symmetry is fully 
broken and the string order parameters 
${\cal{O}}^{\alpha}_{\mbox{\small string}}$ 
($\alpha=z,x$) have finite values, 
while in the N\'eel phase ($\Delta_{c1}<\Delta$) 
the discrete symmetry is partially broken and 
${\cal{O}}^{z}_{\mbox{\small string}}$ 
is non zero but ${\cal{O}}^{x}_{\mbox{\small string}}=0$\cite{KT3}. 
Introducing $\delta$, we expect the dimer phase, in which two $S=1$ spins at 
$j=2i-1$-th and $2i$-th sites ($\delta>0$) make a singlet. 
In this phase, 
the string order parameters ${\cal{O}}^{z,x}_{\mbox{\small string}}$ are zero, 
so that the hidden discrete symmetry is not broken. 
We have obtained the phase diagram with the same topology 
of the quantum Ashkin-Teller model which has the $Z_{2}\times Z_{2}$ symmetry 
explicitly\cite{KNO}. 

But our phase diagram is not consistent with the previous one by 
Singh and Gelfand\cite{Singh}. 
They treated the model(\ref{ham}) by the series expansion and 
showed that there exists a multicritical point of the $XY$, the Haldane, 
the dimer, and the N\'eel phases. 
However, based on the $Z_{2}\times Z_{2}$ symmetry ,
or the phase diagram of the quantum Ashkin-Teller model, 
there should be a 2D Gaussian critical line 
(Haldane-dimer critical line), which contradicts the phase diagram 
obtained by Singh and Gelfand. 

In our previous paper we mainly studied the BKT transitions (XY-Haldane, 
XY-dimer). 
In this paper, we investigate the universality class and 
the critical properties of not only the XY-Haldane and the XY-dimer transitions 
but also of the Gaussian transition between the Haldane and the dimer phases, 
and the 2D Ising transitions of the N\'eel and the dimer phases. 
We clarify the relation between  the hidden $Z_{2}\times Z_{2}$ symmetry 
and Haldane-dimer transition, using twisted boundary conditions. 
We determine the phase boundary and evaluate the critical exponents  
using the exact diagonalization for the finite size systems. 

The organization of this paper is as follows. 
In the next section, we review the quantum Ashkin-Teller model which has 
the $Z_{2}\times Z_{2}$ symmetry explicitly, and as its effective theory 
we consider the sine-Gordon model which describes the information 
for the critical properties of the model(\ref{ham}).
In section 3, we study the critical properties of the $XY$-Haldane and 
the $XY$-dimer (BKT type), the Haldane-dimer (2D Gaussian type), and 
the Haldane-N\'eel and the dimer-N\'eel (2D Ising type) transitions using 
the conformal field theory with the numerical analysis. 
The last section is the conclusion.

\section{Universality class, $Z_{2}\times Z_{2}$ symmetry and $U(1)$ symmetry}
In order to understand the possible phase diagram 
with the $Z_2\times Z_2$ symmetry, 
we review the one dimensional quantum Ashkin-Teller model\cite{Kohmoto}
\begin{eqnarray}
  H_{AT} &=& -\sum_{j} [\sigma_{j}^{z}\sigma_{j+1}^{z} 
                   + \tau_{j}^{z}\tau_{j+1}^{z} 
  + \lambda\sigma_{j}^{z}\sigma_{j+1}^{z} \tau_{j}^{z}\tau_{j+1}^{z}]
  \nonumber \\
     &&  - \beta \sum_{j} [\sigma_{j}^{x} + \tau_{j}^{x} 
      + \lambda\sigma_{j}^{x}\tau_{j}^{x}],
\label{AT}
\end{eqnarray}
where $\sigma_{j}^{z,x}, \tau_{j}^{z,x}$ are Pauli matrices. 
This Hamiltonian is invariant under 
$\sigma^{z}\leftrightarrow -\sigma^{z}$ and  
$\tau^{z}\leftrightarrow -\tau^{z}$ transformations respectively 
($Z_{2}\times Z_{2}$ symmetry). 
For this model(\ref{AT}), 
there is a duality transformation\cite{Kohmoto}
\begin{eqnarray}
  \tilde{\sigma}_{j+1/2}^{x} = \sigma_{j}^{z}\sigma_{j+1}^{z}, &\hspace{5mm}& 
  \tilde{\sigma}_{j+1/2}^{z} = \prod_{i<j}\sigma_{i}^{x},
  \label{dltt}\\
  \tilde{\tau}_{j+1/2}^{x} = \tau_{j}^{z}\tau_{j+1}^{z}, &\hspace{5mm}& 
  \tilde{\tau}_{j+1/2}^{z} = \prod_{i<j}\tau_{i}^{x}.
  \nonumber
\end{eqnarray}
Under this transformation, the Hamiltonian is transformed as
\[
  H_{AT}(\lambda,\beta) = \beta H_{AT}(\lambda,1/\beta),
\]
so at $\beta=1$ the system has the self-duality. 
It is known that there is the 2D Gaussian critical line ($\beta = 1$, 
$-1/\sqrt{2}<\lambda < 1$) of 
continuously varying critical 
exponents, and at one end ($\beta = 1$, $\lambda = 1$), 
this line breaks up into 
two 2D Ising critical lines.
At another end of the Gaussian critical line 
($\beta = 1$, $\lambda = -1/\sqrt{2}$), 
it meets two BKT critical lines 
\cite{BKT}, so there is a massless region or the ``critical fan'' 
in the quantum Ashkin-Teller model.
The Gaussian critical line separates the fully ordered 
and the fully disordered phases about the $Z_{2}\times Z_{2}$ symmetry, 
and the two 2D Ising critical lines are the boundaries of the partially 
broken phase.

According to Kohmoto, den Nijs, and Kadanoff \cite{Kohmoto}, 
the quantum Ashkin-Teller model(\ref{AT}) can be mapped 
to the $S=1/2$ XXZ chain with bond-alternation (\ref{ham}) 
($\Delta=\lambda$, $\delta=(1-\beta)/(1+\beta)$). 
For this model the Gaussian critical line corresponds to
the $\delta=0$ line ($-1/\sqrt{2}<\Delta <1$), and at the
$\Delta=1,\delta=0$ point the bifurcation to the 2D Ising critical
lines occurs.
The universality class of the $\Delta=1,\delta=0$ point is 
of the level-1 $SU(2)$ Wess-Zumino-Witten(WZW) type\cite{AGSZ}.  


The effective  Hamiltonian of the 1D XXZ model with bond-alternation\cite{S} 
(or the quantum Ashkin-Teller model \cite{kdnff}) is given by
\begin{eqnarray}
  H &=&\frac{1}{2\pi}\int dx\left[
    vK(\pi\Pi)^{2}+\frac{v}{K}\left(\frac{\partial\phi}{\partial x} \right)^{2}
      \right] 
\label{dsg} \\
    && +\frac{y_{1}v}{2\pi a^{2}}\int dx\cos\sqrt{2}\phi
        +\frac{y_{2}v}{2\pi a^{2}}\int dx\cos\sqrt{8}\phi,
\nonumber
\end{eqnarray}
where $\Pi$ is the momentum density conjugate to $\phi$,
$
  [\phi(x),\Pi(x^{'})] = i\delta(x-x^{'}),
$
$a$ is the lattice constant and $v$ is the sound velocity. 
The dual field $\theta(x)$ is defined as $\partial_{x}\theta(x) = \pi\Pi(x)$.
Here we make the identification 
$\phi \equiv \phi+\sqrt{2}\pi,\theta \equiv \theta + \sqrt{2}\pi$, 
which means the $U(1)$ symmetry for the $\theta$ field.  
For the free field theory ($y_{1}=y_{2}=0$), 
the scaling dimensions of the operators 
$\exp (\pm im\sqrt{2}\phi)$ and $\exp(\pm in\sqrt{2}\theta)$ are $Km^{2}/2$ and 
$n^{2}/2K$ respectively (where integer variables $m$ and $n$ are magnetic and 
electric charges in the Coulomb gas picture\cite{kdnff-C}).
Since the second term of eq.(\ref{dsg}) is the mass term for the Haldane
gap systems, we have $y_{1}=0$ for the 
half odd integer $S$ XXZ model, and $y_{1}\neq 0$ for the integer $S$ 
XXZ model\cite{S}.

The correspondence between the double sine-Gordon model(\ref{dsg}) 
and the quantum Ashkin-Teller model is as follows\cite{Kohmoto}
\[
  y_{1} \propto \frac{1-\beta}{1+\beta},\hspace{5mm}
  K = \frac{\pi}{\mbox{arccos}(\lambda)}.
\]
Note that the sine-Gordon model(\ref{dsg}) 
is invariant under the transformation
\begin{equation}
  \phi \rightarrow \phi+\frac{\pi}{\sqrt{2}}, \hspace{5mm}
  \theta \rightarrow \theta, \hspace{5mm}
  y_{1} \rightarrow -y_{1}, \hspace{5mm} \mbox{and} \hspace{2mm}
  y_{2} \rightarrow y_{2},
\label{duala}
\end{equation}
which corresponds to the dual transformation(\ref{dltt}) of the quantum Ashkin-Teller 
model.

After the scaling transformation $a\rightarrow e^{dl}a$, we have the 
following renormalization group equations.
\begin{eqnarray}
  \frac{d}{dl}\frac{1}{K} &=& \frac{1}{8}y_{1}^{2} + \frac{1}{2}y_{2}^{2},
    \nonumber \\
  \frac{dy_{1}}{dl} &=& (2-\frac{K}{2})y_{1} - \frac{1}{2}y_{1}y_{2},
\label{kosterlitz} \\
  \frac{dy_{2}}{dl} &=& (2-2K)y_{2} - \frac{1}{4}y_{1}^{2}.
    \nonumber 
\end{eqnarray}
(This form of equations is previously obtained by Kadanoff\cite{kdnff}.) 
In Appendix, we show the derivation of these equations. 
Up to the first order of $y$'s, we see that $y_{1}$ is an irrelevant field 
for $K>4$ and relevant for $K<4$, while $y_{2}$ is irrelevant for $K>1$. 
Thus in the region $K>1$, we can neglect the third term in eq.(\ref{dsg}).
In this case, equations (\ref{kosterlitz}) become
\begin{eqnarray}
  \frac{d}{dl}\frac{1}{K} &=& \frac{1}{8}y_{1}^{2},
    \nonumber \\
  \frac{dy_{1}}{dl} &=& (2-\frac{K}{2})y_{1},
    \nonumber 
\label{recur}
\end{eqnarray}
which are the recursion relations of Kosterlitz\cite{koster}. 
There is a separatrix $32K^{-1}-8\ln K^{-1}-y_{1}^{2}=8+8\ln4$ ($K>4$)
which separates the infrared unstable region from the infrared stable region, 
and on this separatrix, BKT transition occurs. 
On this transition point, logarithmic dependence of finite size spectrum 
appears, and this makes it difficult to extrapolate to the infinite limit 
in the numerical calculation.

The line $y_{1}=0$ ($1<K<4$) is the 2D Gaussian critical line, which separates 
two region. For $1<K<4$ and $y_{1}\neq 0$, $y_{1}$ flows to infinity. 
For $y_{1}>0$, $\langle\phi\rangle$ renormalized to $\pi/\sqrt{2}$ 
as $y_{1}\rightarrow +\infty$ and for $y_{1}<0$, 
$\langle\phi\rangle\rightarrow 0$ as $y_{1}\rightarrow -\infty$.

The equations (\ref{kosterlitz}) are invariant under the transformation 
$y_{1}\rightarrow -y_{1}$, but not invariant under $y_{2}\rightarrow -y_{2}$. 
For the $S=1/2$ bond-alternating XXZ chain(\ref{ham}), $y_{2}>0$ and at the 
$SU(2)$ symmetric point ($\Delta=1$, $\delta=0$) 
the Gaussian line separates to two 2D Ising critical lines. 
There is another possibility other than the Ashkin-Teller multicritical 
structure, depending on the signature of $y_{2}$. 
When $y_{2}$ is negative, the Gaussian line connects to a first order line 
({\it e.g.} $S=1/2$ XXZ chain with staggered field). 

In order to identify the excitations 
with those of the double sine-Gordon model(\ref{dsg}), 
we use the following symmetries. 
The Hamiltonian(\ref{ham}) with the periodic boundary condition 
is invariant under spin rotation around the 
$z$-axis, translation by two-sites 
($\mbox{\boldmath$S$}_{i}\rightarrow \mbox{\boldmath$S$}_{i+2}$), 
space inversion 
($\mbox{\boldmath$S$}_{i}\rightarrow \mbox{\boldmath$S$}_{N-i+1}$) 
and spin reversal 
($S^{z}_{i}\rightarrow -S^{z}_{i}$, $S^{\pm}_{j}\rightarrow -S^{\mp}_{j}$). 
Corresponding eigenvalues are 
$S^z_T \equiv \sum_{i=1}^N S^z_i$, $q=4\pi n/N$ ($n=0,\cdots N/2-1$), 
$P=\pm 1$ and $T = \pm 1$ respectively.
The corresponding symmetry operations 
in the double sine-Gordon model(\ref{dsg}) are as follows.
The operation to the space inversion ($P$) is 
\begin{equation}
  \phi \rightarrow -\phi,\hspace{5mm}
  \theta \rightarrow \theta +\pi/\sqrt{2},\hspace{5mm} x \rightarrow -x,
\end{equation}
and the operation to the spin reversal ($T$) is
\begin{equation}
  \phi \rightarrow -\phi,\hspace{5mm}
  \theta \rightarrow -\theta+\pi/\sqrt{2}.
\end{equation}
The correspondence of these symmetry operations is summarized in Table 1.

Assuming conformal invariance, 
the scaling dimension $x_n$ is related to the energy gap of the finite 
size system with periodic boundary conditions \cite{Cardy} 
\begin{equation}
  x_{n} = \frac{L}{2\pi v}\left(E_{n}(L)-E_{g}(L)\right),
\label{ca}
\end{equation}
where $L$ is the system size, $E_{g}(L)$ is the ground state energy, 
and $v$ is the velocity of the system. 
And the leading finite size correction of the ground state energy is 
\cite{Blote-C,Affleck86}
\begin{equation}
  E_{g}(L) = \epsilon_{g}L - \frac{\pi v c}{6L},
\label{ano}
\end{equation}
where $c$ is the conformal anomaly number. 

In the following section, we calculate these values from small size data 
of the exact diagonalization with the Lanczos method. 

\section{Phase transitions and Numerical results}
In this section, we determine the critical points and the universality class. 
From the hidden $Z_{2}\times Z_{2}$ symmetry and the phase structure of the 
quantum Ashkin-Teller model, we expect that 
the XY-Haldane and the XY-dimer transitions are of the BKT type, 
the Haldane-dimer transition is of the Gaussian type, 
and the Haldane-N\'eel and the dimer-N\'eel 
transitions belong to the 2D Ising universality class. 
With the operator structure of the sine-Gordon model(\ref{dsg}) 
and the renormalization group analysis, 
we determined the transition points numerically. 
To check the universality class, we calculate numerically the conformal anomaly number $c$ 
and several relations between scaling dimensions. 

\subsection{Berezinskii-Kosterlitz-Thouless (XY-Haldane and XY-dimer) transitions} 
First we consider the XY-dimer and the XY-Haldane phase
transitions. 

For $\delta=0$ case ($XY$-Haldane), several numerical studies have been done. 
For the transition point, Botet and Julian\cite{BJ} estimated 
$\Delta_{c}\sim 0.1$, 
Sakai and Takahashi\cite{ST} $\Delta_{c}=-0.01\pm 0.03$, 
and Yajima and Takahashi\cite{YT} $\Delta_{c}=0.069\pm 0.003$. 
They expected that the transition point is 
exactly $\Delta=0$ (see also \cite{AM,AlH,Malvezzi}). 
In our previous paper, we found  that the transition point between 
the Haldane and the $XY$ phases is exactly $\Delta=0$ 
in the numerical accuracy\cite{KNO}. 

The effective Hamiltonian for this transition is described by the 
sine-Gordon model,
\begin{eqnarray}
  H &=&\frac{1}{2\pi}\int dx\left[
    vK(\pi\Pi)^{2}+\frac{v}{K}\left(\frac{\partial\phi}{\partial x} \right)^{2}
      \right]  \nonumber \\
    && +\frac{y_{1}v}{2\pi a^{2}}\int dx\cos\sqrt{2}\phi. 
\label{con}
\end{eqnarray}
Here we neglect the third term of eq.(\ref{dsg}) which is irrelevant 
for the BKT transition.
For this model, there is a special operator, that is the marginal operator 
\begin{equation}
  {\cal{M}} = -a^{2}K(\pi\Pi)^{2}
    +a^{2}\frac{1}{K}\left(\frac{\partial\phi}{\partial x}
  \right)^{2},
\end{equation}
(this is the Lagrangian density itself for the free field theory) 
which changes the scaling dimension of scaling operators continuously. 
For the free field theory, 
the scaling dimension of $\sqrt{2}\cos\sqrt{2}\phi$ 
in the second term of eq.(\ref{con}) is $K/2$, and the BKT transition occurs 
when this dimension becomes $2$, {\it i.e.}, the renormalized $K$ is $4$.

Besides the operators ${\cal{M}}$ and $\sqrt{2}\cos\sqrt{2}\phi$, 
$\sqrt{2}\sin\sqrt{2}\phi$ and $\exp(\mp i4\sqrt{2}\theta)$ are also 
marginal at the BKT transition point. 
Recently Nomura \cite{Nomura95} studied the structure of these
operators near the transition point. 
He found that the $\sqrt{2}\cos\sqrt{2}\phi$ and the marginal operator ${\cal{M}}$
are hybridized in the course of the renormalization, to satisfy the 
orthogonality condition of scaling operators. 
We set the scaling dimensions of $\sqrt{2}\cos\sqrt{2}\phi$-like, 
$\sqrt{2}\sin\sqrt{2}\phi$, $\exp(\mp i4\sqrt{2}\theta)$ and marginal-like operators 
as $x_{1},x_{2},x_{3}$ and $x_{0}$.
We denote $K$ and $y_{1}$ as $K=4+2y_{0}$ and $y_{1}=\pm y_{0}(1+t)$ 
near the BKT critical point ($y_{1} = \pm y_{0}$). 
Since the scaling dimension is related to the excitation energy as 
eq.(\ref{ca}), 
Nomura found the finite size behaviors of scaling dimensions 
up to the first order of $y_{0}$ and $y_{1}$ as
\begin{eqnarray}
  x_{1}(l) &\equiv& \frac{L\Delta E_{1}(L)}{2\pi v} 
  = 2+2y_{0}(l)(1+\frac{2}{3}t),
  \nonumber \\
  x_{2}(l) &\equiv& \frac{L\Delta E_{2}(L)}{2\pi v} = 2+y_{0}(l),
\label{corr}  \\
  x_{3}(l) &\equiv& \frac{L\Delta E_{3}(L)}{2\pi v} = 2-y_{0}(l),
  \nonumber \\ 
  x_{0}(l) &\equiv& \frac{L\Delta E_{0}(L)}{2\pi v} 
  = 2-y_{0}(l)(1+\frac{4}{3}t),
  \nonumber
\end{eqnarray}
where we define $l=\log L$. 
From the recursion relations of Kosterlitz(\ref{recur}) for $t=0$, 
we have $y_{0}(l) = 1/\log (L/L_{0})$ ($L_{0}$ is a constant), so logarithmic 
corrections appear. 
At the transition point the scaling dimensions $x_{0}$
and $x_{3}$ cross linearly, and 
this means that at the critical point there exists a degeneracy of excited 
states. 
So we can determine the BKT transition point by the level crossing of 
these excitations, and we can also eliminate the logarithmic corrections by 
the appropriate average of the excitations. 
From table.1, the first $S^z_T=0, q=0, P=T=1$ [$x_{0}$]
and the $S^z_T=\pm 4, q=0, P=1$ [$x_{3}$] excitations 
degenerate on the BKT line. 
Note that in the $S^z_T=\pm 1, P=-1$ subspace the lowest excitations 
correspond to the operators $\exp( \mp i\sqrt{2}\theta)$ 
whose scaling dimension is $x_{4}=1/8$ ($\eta=1/4$)
at the BKT transition point.

In numerical calculations, we study finite rings of $N$ sites 
($N = 8,10,12,14,16$) with periodic boundary conditions.
We determine the velocity $v$ by the current field ($x=1$, $q=2\pi/L$),
\[
    v = \lim_{L\rightarrow\infty}\frac{\Delta E(L,k=2\pi/L)}{2\pi /L},
\]
where $L$ is the system size and $L=N/2$ in the considering case. 
To obtain this value, we extrapolated the value of the finite size system as 
$v(L) = v + \mbox{const}/L^{2}$. 
The leading finite size corrections 
of eq.(\ref{ca}) come from the operators 
$L_{-2}\bar{L}_{-2}${\bf 1} and 
$((L_{-2})^{2}+(\bar{L}_{-2})^{2})${\bf 1}\cite{Cardy,Cardy86,Reinike} 
which are the descendant fields of the identity operator. 
The scaling dimensions of these operators 
are $4$, and the leading correction term of the scaling dimension(\ref{ca}) 
behaves as $1/L^{2}$. 
(Here we neglect the corrections from third term of eq.(\ref{dsg}) because 
the scaling dimension of $\sqrt{2}\cos\sqrt{8}$ is $2K$, bigger than $4$.)
In Fig.2(a) we show the excitation energies for $N=16,\Delta=-0.5$.
We determine the XY-dimer phase boundary by 
the crossing point of excitations $S^z_T = 0, P=T=1$ 
and $S^z_T=\pm 4, P=1$. 
Figure 2(b) shows the excitation energies for $N=16$, $\Delta=0$. 
In this case one of the states ($S^z_T=0$, $q=0$, $P=T=1$) 
and the eigenstate ($S^z_T=\pm4$, $q=0$, $P=1$) are 
exactly degenerate on the whole line, 
thus in our previous paper 
we have concluded that the XY-Haldane phase boundary is on exactly $\Delta=0$.
We determine the multi-critical point of the XY, the dimer, and the Haldane 
phases by the crossing point of excitations $S^{z}_{T}=\pm4$, $P=1$[$x_{3}$] 
and $S^{z}_{T}=0$, $P=T=-1$[$x_{2}$] on the $\Delta=0$ line. 
This point corresponds to 
$K=4$ and $y_{1}=0$ in the continuum Hamiltonian (\ref{con}).
Figure 3 shows the size dependence of this multi-critical point, 
and we have estimated the multi-critical point 
as $\Delta=0, \delta=0.230\pm 0.001$.  

In our previous paper, we have presented 
the size dependence of $(x_{0}+x_{2})/2$ 
and $(2x_{0}+x_{1})/3$ to verify that the transition is of the 
BKT type. From eqs.(\ref{corr}), these values eliminate the logarithmic 
corrections and we have seen that these converge to 2 
as the system size increases. 
In Fig.4 we show the extrapolated values along the BKT critical line. 
From eqs.(\ref{corr}), the average $(3x_{0}+x_{1}+x_{2})/5$ also eliminates 
the logarithmic corrections, where the coefficient $3,1,1$ is 
the degeneracy at the critical point. 
Figure 5 shows the size dependence of it, 
where $1/N^{2}$ behavior comes 
from corrections of the irrelevant operators 
$L_{-2}\bar{L}_{-2}${\bf 1} and 
$((L_{-2})^{2}+(\bar{L}_{-2})^{2})${\bf 1}. 
In Fig.6 we show the extrapolated values along the BKT critical line. 
We have estimated the conformal anomaly number using eq.(\ref{ano}), as 
$c=0.998$ for $\Delta=-0.5$, $\delta=0.583$ 
and $c=1.000$ for $\Delta=0$, $\delta=0$, also considering the finite size 
correction of the $x=4$ irrelevant field. 
Hence we conclude that the transitions are of the BKT-type. 

Next, we show the asymptotic behavior near the BKT critical point. 
From eqs.(\ref{corr}), we can remove the $t$ linear terms as
\begin{equation}
  \frac{1}{3}[x_{0}(l)+x_{1}(l)+x_{3}(l)]
    = 2+{\cal{O}}(y_{0}^{2}).
\end{equation}
In Fig.7, we show this value near the critical points 
$\Delta = -0.5$, $\delta = 0.583$ and $\Delta = 0$, $\delta=0$. 
The linear components of $t$ are almost absent. 

Lastly, we consider the $XY$ region where the energy gap is zero and the spin 
correlations decay in the power law. 
The scaling dimensions of the operators 
$\exp(\mp i\sqrt{2}\theta)$ [$x_{4}$] and 
$\exp(\mp i2\sqrt{2}\theta)$ [$x_{5}]$ are 
$x_{4} = 1/2K$ and $x_{5}=2/K$, 
so the ratio of these values is $x_{5}/x_{4}=4$. 
These operators correspond to the lowest states 
($S^{z}_{T}=\pm1$, $q=0$, $P=-1$) 
and ($S^{z}_{T}=\pm2$, $q=0$, $P=1$) respectively. 
On the BKT transition point, we have 
\begin{eqnarray}
  x_{4}(l) &=& \frac{1}{8}-\frac{y_{0}}{16}, \nonumber \\
  x_{5}(l) &=& \frac{1}{2}-\frac{y_{0}}{4}, 
\label{nonzero}
\end{eqnarray}
so this ratio is correct up to the first order 
of the logarithmic correction $y_{0}(l)=1/\log(L/L_{0})$. 
Figure 8 shows the extrapolated values 
of this ratio for $\Delta=-0.5$. The evaluated values are almost $4$, except 
near the BKT transition point. The discrepancy near the transition point 
comes from the higher order logarithmic corrections 
[${\cal{O}}(1/(\ln L)^{2})$]. 

In the XY region, 
the product $x_{2}\cdot x_{3}$ is $4$ 
up to ${\cal{O}}(1/(\ln L)^{2})$ (from eqs.(\ref{corr})) 
near the transition point, 
since the field $x_{2}$ is free from the effect of the hybridization. 
(Contrary to this, the product $x_{1}\cdot x_{3}$ is affected 
from the bybridization of $x_{0}$ and $x_{1}$, so there remains 
${cal{O}}(1/\ln L)$ correction.) 
In Fig.9, this product value is presented and we can see the expected value. 

In Fig.10, we show the scaling dimension $x_{0}$ of 
the marginal operator ${\cal{M}}$ of the $N=16$ system. 
On the transition point the deviation from $2$ is more than 10 \%. 
The deviation of $x_{0}$ from $2$ comes from the hybridization with the operator 
$\sqrt{2}\cos\sqrt{2}\phi$. 
Near the transition point 
we can eliminate this effect of the hybridization taking the combination 
$x_{0}+x_{1}-x_{2}=2$, and so this value eliminate 
the logarithmic correction up to the first order of $y_{0}$. 
In Fig.10 we show the extrapolated value of $x_{0}(l)+x_{1}(l)-x_{2}(l)$, 
and the discrepancy decreases in a few percent.

\subsection{Gaussian (Haldane-dimer) transition}
Next, let us consider the Haldane-dimer transition. 
In the dimer phase, the system has no symmetry breaking, whereas 
in the Haldane phase, the system exhibits 
the hidden $Z_{2}\times Z_{2}$ symmetry breaking.
For this transition, the operator $\sqrt{2}\cos\sqrt{2}\phi$ 
in eq.(\ref{dsg}) is relevant($4>K>1$) and the Gaussian transition occurs at 
$y_{1}=0$. 

\subsubsection{Determination of the transition point}
In the content of the VBS picture, 
for even site systems with periodic boundary conditions, 
the Haldane gap state of $S=1$ systems can be written as\cite{AAH} 
\begin{equation}
(a^{\dagger}_{N}b^{\dagger}_{1}-b^{\dagger}_{N}a^{\dagger}_{1})\prod_{j=1}^{N-1}
(a^{\dagger}_{j}b^{\dagger}_{j+1}-b^{\dagger}_{j}a^{\dagger}_{j+1}) |0\rangle,
\end{equation}
and the dimer state can be
\begin{equation}
  \prod_{k=1}^{N/2}
  (a^{\dagger}_{2k-1}b^{\dagger}_{2k}-b^{\dagger}_{2k-1}a^{\dagger}_{2k})^{2}
  |0\rangle \hspace{5mm}(\delta >0),
\end{equation}
or
\[
  (a^{\dagger}_{N}b^{\dagger}_{1}-b^{\dagger}_{N}a^{\dagger}_{1})^{2}
  \prod_{k=1}^{N/2-1}
  (a^{\dagger}_{2k}b^{\dagger}_{2k+1}-b^{\dagger}_{2k}a^{\dagger}_{2k+1})^{2}
  |0\rangle \hspace{5mm}(\delta<0),
\]
where we describe the spin state by the Schwinger bosons, 
that is, $a^{\dagger}_{j}$ ($b^{\dagger}_{j}$) creates 
the $S=1/2$ $\uparrow$ ($\downarrow$) spin at the $j$-th site.
These two states have the same symmetries which conserve the Hamiltonian, 
especially the space inversion $j\rightarrow N-1+j$ ($P=1$) 
and the spin reversal $a^{\dagger}\leftrightarrow b^{\dagger}$ ($T=1$).
For the periodic boundary conditions, the ground state is 
singlet in both the Haldane and the dimer phases. 
So we cannot expect the simple finite size scaling behaviors near the 
critical points. 

But the situation is different for another boundary condition, {\it i.e.}, 
the twisted boundary condition $S^{x}_{N+1}= -S^{x}_{1}$, 
$S^{y}_{N+1}= -S^{y}_{1}$, $S^{z}_{N+1}= S^{z}_{1}$. 
For this boundary condition, 
the Haldane gap state is written as 
\begin{equation}
(a^{\dagger}_{N}b^{\dagger}_{1}+b^{\dagger}_{N}a^{\dagger}_{1})\prod_{j=1}^{N-1}
(a^{\dagger}_{j}b^{\dagger}_{j+1}-b^{\dagger}_{j}a^{\dagger}_{j+1}) |0\rangle,
\end{equation}
while the dimer state can be
\begin{equation}
  \prod_{k=1}^{N/2}
  (a^{\dagger}_{2k-1}b^{\dagger}_{2k}-b^{\dagger}_{2k-1}a^{\dagger}_{2k})^{2}
  |0\rangle \hspace{5mm}(\delta>0), 
\end{equation}
or
\[
  (a^{\dagger}_{N}b^{\dagger}_{1}+b^{\dagger}_{N}a^{\dagger}_{1})^{2}
  \prod_{k=1}^{N/2-1}
  (a^{\dagger}_{2k}b^{\dagger}_{2k+1}-b^{\dagger}_{2k}a^{\dagger}_{2k+1})^{2}
  |0\rangle \hspace{5mm}(\delta<0).
\]
The Haldane state has $P=-1$, $T=-1$, but 
the dimer state has $P=1$, $T=1$, 
so these two states have the different $P$ and $T$ eigenvalues, 
and the energy eigenvalues cross at the critical point.

For the sine-Gordon model, we can explain this as follows. 
First, let us consider the following 1-D Hamiltonian 
with periodic boundary condition
\begin{equation}
  H = H_{0}+\frac{\lambda}{2\pi}\int_{0}^{L} dx{\cal{O}}_{1},
\end{equation}
where $H_{0}$ is the fixed point Hamiltonian, 
$L$ is the system length (setting the ultraviolet cutoff as $1$), 
and ${\cal{O}}_{1}$ is a scaling operator 
whose scaling dimension is $x_{1}$ (${\cal{O}}^{\dagger}_{1}={\cal{O}}_{1})$.
According to Cardy\cite{Cardy}, the following size dependence of excitation 
energy up to the first order perturbation is obtained
\begin{equation}
  \Delta E_{n} = \frac{2\pi v}{L}\left(
    x_{n}+\lambda C_{n1n}\left(\frac{2\pi}{L}\right)^{x_{1}-2}
    +\cdots \right),
\label{rel}
\end{equation}
where $x_{n}$ is the scaling dimension 
of the operator ${\cal{O}}_{n}$, 
and $C_{n1n}$ is the operator product expansion (OPE) 
coefficient of the operators ${\cal{O}}_{n}$ and ${\cal{O}}_{1}$.
We assume that the operator ${\cal{O}}_{1}$ is a relevant one ($x_{1}<2$). 
Up to the first order perturbation theory, we find that 
at the critical point $\lambda=0$ 
the scaled gap $L\Delta E_{n}$ does not 
depend on the system size besides 
finite size corrections of irrelevant operators.
But if the OPE coefficient $C_{n1n}$ is zero, 
the first order perturbation theory is not sufficient 
and we must consider the second order term of $\lambda$ in eq.(\ref{rel}). 
In such a case the scaled gap $L\Delta E_{n}$ has some extremum at the 
critical point.

In the sine-Gordon model, the relevant operator is 
${\cal{O}}_{1}=\sqrt{2}\cos\sqrt{2}\phi$, 
and there is no operator ${\cal{O}}_{n}$ 
which has nonzero value of $C_{n1n}$. 
It is related with the charge neutrality conditions 
in the Coulomb gas picture\cite{kdnff-C}. 
(Note that the operators $e^{\pm i\phi/\sqrt{2}}$ is not allowed by the identification $\phi\equiv\phi+\sqrt{2}\pi$.) 
Thus the OPE coefficient in eq.(\ref{rel}) is zero, 
and we cannot use the usual finite size scaling method.
In other words, the Haldane gap systems with the periodic boundary condition
does not have the local ``order parameter" 
which has non-zero value of OPE coefficient with the relevant operator 
in the Hamiltonian\cite{Nijs}.

Second, let us see what happens for the twisted boundary 
condition 
$S^{x}_{N+1}\pm iS^{y}_{N+1} = e^{\pm i\Phi}(S^{x}_{1}\pm iS^{y}_{1})$, 
$S^{z}_{N+1}=S^{z}_{1}$. 
It has been known that 
for the primary operator $\exp(i\sqrt{2}m\phi+i\sqrt{2}n\theta)$, 
the effect of the boundary 
condition $\Phi=\pi$ is to shift the magnetic charge $m$ by one half\cite{ABB} 
$m\rightarrow m-1/2$, $n\rightarrow n$. 
If the half magnetic charges exist, we have the following size dependence 
of the scaling dimensions of the half magnetic charge operators 
$\sqrt{2}\cos\phi/\sqrt{2}$ [$x_{6}$] and 
$\sqrt{2}\sin\phi/\sqrt{2}$ [$x_{7}$] as\cite{KTZB}
\begin{eqnarray}
  x_{6}(L) &=& \frac{L(E_{6}(L,\Phi=\pi)-E_{g}(L,\Phi=0))}{2\pi v}\nonumber \\ 
    &=& \frac{K}{8} + \frac{y_{1}}{2}
      \left(\frac{2\pi}{L}\right)^{K/2-2} + \cdots , \nonumber \\
  x_{7}(L) &=& \frac{L(E_{7}(L,\Phi=\pi)-E_{g}(L,\Phi=0))}{2\pi v} 
\label{cross} \\
    &=& \frac{K}{8} - \frac{y_{1}}{2}
      \left(\frac{2\pi}{L}\right)^{K/2-2} + \cdots. 
\nonumber
\end{eqnarray}
We see that these two energy eigenvalues cross linearly at $y_{1} = 0$. 

Note that the sine-Gordon model(\ref{dsg}) 
is invariant under the transformation(\ref{duala}). 
The half magnetic charge 
operator $\sqrt{2}\cos\phi/\sqrt{2}$ ($\sqrt{2}\sin\phi/\sqrt{2}$) corresponds 
to the operator $P=\sigma^{z}\tau^{z}$ 
($\tilde{P}=\tilde{\sigma}^{z}\tilde{\tau}^{z}$)\cite{kdnff-brwn} 
of the quantum Ashkin-Teller model.
Under this transformation, the half magnetic charge operators $\sqrt{2}\cos\phi/\sqrt{2}$ and 
$\sqrt{2}\sin\phi/\sqrt{2}$ are transformed as
\begin{eqnarray}
  \sqrt{2}\cos\frac{1}{\sqrt{2}}\phi &\rightarrow& 
   -\sqrt{2}\sin\frac{1}{\sqrt{2}}\phi, \nonumber \\
  \sqrt{2}\sin\frac{1}{\sqrt{2}}\phi &\rightarrow& 
    \sqrt{2}\cos\frac{1}{\sqrt{2}}\phi.
\end{eqnarray}
Thus at the point $y_{1} = 0$, the system has the self-duality 
in the quantum Ashkin-Teller language. 
By the invariance of eq.(\ref{duala}), the third term of eq(\ref{dsg}) 
does not affect the determination of the crossing point for eq.(\ref{cross}). 

Numerically, with $\Phi=\pi$ boundary condition 
we determine the phase boundary of the Haldane and dimer phases 
for $N=8,10,12,14,16$ systems. 
In Fig.11, we show two low lying energies of the subspace $\sum S^{z} = 0$ 
with the boundary condition $S^{x}_{N+1}=-S^{x}_{1}$, 
$S^{y}_{N+1}=-S^{y}_{1}$, 
$S^{z}_{N+1}=S^{z}_{1}$ for $\Delta = 0$, $0.5$ and $1$, 
which correspond to $E_{6}(L,\Phi=\pi)-E_{g}(L,\Phi=0)$ 
and $E_{7}(L,\Phi=\pi)-E_{g}(L,\Phi=0)$.
From figures of $\Delta=0.5$, $1.0$, we see that in the Haldane phase $P=T=-1$ state 
is lower than $P=T=1$ state but in the dimer phase the relation reverses 
as expected. 

However, contrary to the $S=1/2$ case, since the self-duality at $y_{1}=0$ 
is not exact for $S=1$ bond-alternating system, the crossing point 
of $E_{6}$ and $E_{7}$ has a size dependence as ${\cal{O}}(N^{-2})$ 
(or ${\cal{O}}(N^{-4})$) (see appendix B). 
Figure 3($\Delta=0$) and Figure 12($\Delta=0.5$) show the size dependence 
of the crossing points. 
In fact, assuming the size dependence of the crossing point
$\delta_{c}(N)= \delta_{c}(\infty)+aN^{-b}$, we can check the power $b$, using 
the relation
\begin{equation}
   \frac{
   \ln\left(\frac{\delta_{c}(N+2)-\delta_{c}(N)}{\delta_{c}(N)-\delta_{c}(N-2)}
   \right)}{
   \ln\left(\frac{N+1}{N-1}\right)}
   = -b -1 +{\cal{O}}\left(\frac{1}{N^{2}}\right). 
\label{expnntg}
\end{equation}
Figure.13 shows the exponet evaluated with this method. 
Hereafter we extrapolate the critical point as 
$\delta_{c}(N)= \delta_{c}(\infty) +  aN^{-2} + bN^{-4}$.

\subsubsection{Check of the universality class}
In Fig.14, we show the value $K$ estimated from $x_{1}$,$x_{2}$, $x_{4}$, 
and $x_{6}$ of the $N=16$ system. These values are almost consistent 
except near $\Delta=1$ where the effect of the logarithmic corrections of 
the level-1 $SU(2)$ WZW model\cite{AGSZ} appears.

We estimated the conformal anomaly number at the critical point $\Delta=0.5$, 
$\delta=0.2524$ as $c=0.998$. 
In Table 2, we show some extrapolated scaling dimensions for $\Delta=0.5$. 
The estimation of $K$ is consistent in each others.

For the Gaussian transitions the leading finite size corrections 
of eq.(\ref{ca}) come from the operators 
$L_{-2}\bar{L}_{-2}${\bf 1} and 
$((L_{-2})^{2}+(\bar{L}_{-2})^{2})${\bf 1} or 
$\sqrt{2}\cos\sqrt{8}\phi$, of which the third operator 
is included in the third term of eq.(\ref{dsg}). 
The scaling dimensions of these operators 
are $4$ and $2K$ ($1<K<4$) respectively. 
By the third term of eq.(\ref{dsg}), 
at $y_{1}=0$ only the scaling dimensions $x_{1}$ and $x_{2}$ 
are corrected as 
\begin{eqnarray}
  x_{1}(L) &=& \frac{K}{2}+\frac{y_{2}}{2}\left(\frac{1}{L}\right)^{2K-2}
  +\cdots,  \nonumber \\
  x_{2}(L) &=& \frac{K}{2}-\frac{y_{2}}{2}\left(\frac{1}{L}\right)^{2K-2}
  +\cdots.  \nonumber
\end{eqnarray}
We can eliminate this correction by summing these two dimensions. 
To improve the precision of the 1-parameter scaling, hereafter 
we try to eliminate the effect of the $y_{1}$, $y_{2}$ fields, 
at least first order. 

We show the extrapolated values of $x_{0}$ on the Haldane-dimer transition lines
in Fig.15. The discrepancy is about $30$ \% at $\Delta=1$. 
Figure 16 shows the extrapolated values of 
\[
  x_{4}\cdot (x_{6}+x_{7}) = \frac{1}{2K}\cdot (\frac{K}{8}+\frac{K}{8}), 
\]
and the obtained values are consistent with $1/8$ except near $\Delta=1$, 
where the discrepancy (about 15 \%) 
comes from the logarithmic corrections of higher orders. 
In Fig.17, we present the ratio $x_{4}$ and $x_{5}$. 
Figure 18 shows the extrapolated values of $(x_{1}+x_{2})\cdot x_{4}$ which 
should be $4$. The summation of $x_{1}$ and $x_{2}$ eliminate the correction 
from the third term of eq.(\ref{dsg}). 
In Fig.19, we present the ratio $(x_{1}+x_{2})/(x_{6}+x_{7})$ where 
the numerator is the value with the periodic boundary condition and 
the denominator is with the $\Phi=\pi$ twisted boundary condition. 

Last of this subsection, we remark that 
the method to determine the transition point can also 
apply to determine the Gaussian fixed point ($y_{1}=0$) in the XY phase. 
In Fig.1, we also show the Gaussian fixed line.
We find that the line ends at the point $\Delta=-1$, $\delta=0$. 
Near this point, 
the Gaussian line behaves as $\delta\propto (1+\Delta)^{1/4}$ (see Fig.20).

\subsection{Isotropic case}
The isotropic case $\Delta=1$ has been studied in relation with the 
Affleck-Haldane prediction. Affleck and Haldane\cite{AFF,AFFH} considered the $\Delta=1$ 
case for the arbitrary $S$ case, mapping onto the nonlinear $O(3)$ 
$\sigma$-model. They showed that the topological angle $\theta$ is 
given by $\theta=2\pi S(1-\delta)$ and the system should be massless 
when $\theta/2\pi$ is half odd integer. 
Numerically several authors estimated 
the transition point $\delta_{c}$. 
For $S=1$ case, they concluded 
$\delta_{c}\simeq 0.25\pm 0.01$ and $c=1$\cite{KT2,Y,Te}.
Totsuka {\it et. al.}\cite{Te} extensively studied this case, and 
obtained the critical dimensions up to the 
logarithmic corrections and identified the universality class as the 
$SU(2)$ level-1 Wess-Zumino-Witten model. 

For $y_{1}=0$, the RG equations are given by 
\begin{eqnarray}
  \frac{d}{dl}\frac{1}{K} &=& \frac{1}{2}y_{2}^{2}, \nonumber \\
  \frac{dy_{2}}{dl} &=& 2(1-K)y_{2}.
\end{eqnarray}
Denoting $K=1+y_{0}/2$ near $K=1$, 
we have
\begin{eqnarray}
  \frac{dy_{0}}{dl} &=& -y_{2}^{2} \nonumber, \\
  \frac{dy_{2}}{dl} &=& -y_{0}y_{2}. \nonumber
\end{eqnarray}
The WZW point is $y_{0}=y_{2}$. 
According to Giamarchi and Schulz\cite{gmrch-schlz}, 
and Nomura\cite{Nomura95}, we obtain the following 
scaling dimensions
\begin{eqnarray}
  x_{0}(l) &=& 2-y_{0}(l)\left(1+\frac{4}{3}t\right), \nonumber \\
  x_{1}(l) &=& \frac{1}{2}+\frac{3}{4}y_{0}(l)\left(1+\frac{2}{3}t\right), 
  \nonumber \\
  x_{2}(l) &=& \frac{1}{2}-\frac{1}{4}y_{0}(l)(1+2t),
  \label{rgi} \\
  x_{4}(l) &=& \frac{1}{2}-\frac{1}{4}y_{0}(l), \nonumber \\
  x_{5}(l) &=& 2-y_{0}(l), \nonumber \\
  x_{8}(l) &=& 2+2y_{0}(l)\left( 1+\frac{2}{3}t\right), \nonumber \\
  x_{9}(l) &=& 2+y_{0}(l), \nonumber
\end{eqnarray}
where $t$ is defined as $y_{2} = y_{0}(1+t)$, 
and we define the scaling dimensions 
of $\sqrt{2}\cos\sqrt{8}\phi$ and $\sqrt{2}\sin\sqrt{8}\phi$ 
as $x_{8}$ and $x_{9}$ respectively. 
At $t=0$ we have $y_{0}(l) = 1/\log(L/L_{0})$ and 
the eqs.(\ref{rgi}) are consistent 
with the non-Abelian bosonization theory\cite{AGSZ}.

From these equations, we see that $(x_{1}(l)+3x_{4}(l))/4$, 
$(x_{0}(l)+x_{9}(l))/2$, $(2x_{0}(l)+x_{8}(l))/3$, and 
$(5x_{0}+3x_{9}+x_{8})/9$ 
do not depend on the logarithmic corrections 
at the Wess-Zumino-Witten point. 

We obtained the transition point with the same method 
in the previous subsection. The obtained value is $\delta_{c}=0.2598$. 
The conformal anomaly number is estimated as $c=0.990$. 
The extrapolated value of $(x_{1}(l)+3x_{4}(l))/4$ is $0.495$. 
and the extrapolated values of $(x_{0}(l)+x_{9}(l))/2$, 
$(2x_{0}(l)+x_{8}(l))/3$, and $(5x_{0}+3x_{9}+x_{8})/9$ 
are $1.982$, $2.026$, and $1.997$ respectively. 

\subsection{2D Ising (Haldane-N\'eel and dimer-N\'eel) transition}
Lastly, we study the N\'eel-Haldane and the N\'eel-dimer transitions. 
Because in the N\'eel phase, the hidden $Z_{2}\times Z_{2}$ symmetry is partly 
breaking, we see that the broken symmetry of these transitions is 
the $Z_{2}$ symmetry.

On the N\'eel-Haldane transition ($\Delta=1.17\pm0.02$, $\delta=0$), 
Nomura\cite{KN} calculated the critical exponents as $\nu=0.98\pm0.007$, 
$\beta=0.126\pm0.007$,$\eta=0.253\pm0.002$, 
with a (large cluster decomposition) Monte Carlo method. 
Later Sakai and Takahashi\cite{ST} reexamined this transition 
using the phenomenological renormalization group method 
and finite-size scaling technique, and obtained $\nu=1.02\pm0.05$, 
$\gamma/\nu=1.76\pm0.01$, and $\eta=0.23\pm0.01$. These results are 
consistent with the 2D Ising universality class 
($\nu=1$, $\beta=1/8$, $\eta=1/4$, $\gamma=7/4$). 
So here we consider the N\'eel-dimer transition and check 
its universality class.

In $c<1$ conformal field theory, the current field 
does not exist, so we cannot determine the velocity by the same method 
as the $c=1$ case.

Assuming the 2D Ising transition, we expect that 
the first excitation state has the $S^{z}_{T}=0$, $q=0$, $P=T=-1$ symmetry, 
and this corresponds to the order parameter $\sigma$ of the Ising model,  
whose scaling dimension is $1/8$. 
The scaling dimension of the level-1 descendant field $\hat{L}_{-1}\sigma$ 
is $1+1/8$, 
so we can determine the velocity of the system by 
\begin{equation}
  v = \frac{E_{\hat{L}^{-1}\sigma}-E_{\sigma}}{2\pi L}.
\label{vel}
\end{equation}
Using this velocity we can determine the conformal anomaly number $c$ and 
scaling dimensions of several excitations. 

Numerically we determined the phase boundary 
by the phenomenological renormalization group  method, 
that is, evaluating the crossing point of the scaled gaps 
$N\Delta E(N)$ and $(N+2)\Delta E(N+2)$ 
for $N=6,8,10,12,14$ systems with periodic boundary conditions. 
Using the velocity(\ref{vel}), 
we estimate the conformal anomaly number as $c=0.4997$ and 
the scaling dimensions as $x_{\sigma}=0.126$ and $x_{\epsilon}=1.001$, 
for $\Delta=2.0$, $\delta=0.683$. 
In figure 21, we show low lying scaling dimensions 
of the dimer-N\'eel transition point, 
where circle shows $S^{z}_{T}=0$, $P=T=1$, and cross shows 
$S^{z}_{T}=0$, $P=T=-1$ states. 
These values are consistent with the prediction from the Ising universality.

\section{Conclusion}
We discussed the critical properties of 
the $S=1$ bond-alternating XXZ spin chains. 
The expected phases are the ferromagnetic, the $XY$, the Haldane gap, 
the dimer, the N\'eel phases. 
First, we considered the universality class based on 
the hidden $Z_{2}\times Z_{2}$ symmetry. 
Then we have taken the analogy from the quantum Ashkin-Teller model 
which has a $Z_{2}\times Z_{2}$ symmetry explicitly. 
The effective model of the quantum Ashkin-Teller model is the double 
sine-Gordon model, and using the information of it, we determined 
the XY-Haldane, the XY-dimer and the Haldane-dimer phase boundaries 
numerically. 
For the XY-Haldane and the XY-dimer (BKT) transition, 
the renormalization group aspect is important and it is difficult 
to apply the simple finite size scaling method due to the logarithmic 
corrections. While for the Gaussian transition between the Haldane and 
the dimer phases, we adapted the twisted boundary condition 
to obtain the preferable operator structure and determined 
the transition point by a level crossing. 
This method can also be applied to determine the Gaussian fixed line 
in the XY phase. 
We found that the hidden discrete symmetry is crucial for 
the phase transition and the topology of the phase diagram. 
We also identified the universality class numerically using the conformal 
field theory and eliminating the correction of the (marginal) irrelevant fields. 


It is interesting to consider the arbitrary spin case. Oshikawa\cite{OSHI} studied 
the arbitrary $S$ Heisenberg chain with bond-alternation in the VBS picture.
It was pointed out that the successive dimerization 
transition occurs in the Affleck-Haldane prediction\cite{Guo}.
Oshikawa calculated the string order parameter, and concluded that 
for the successive dimerization, the breakdown 
of the hidden $Z_{2}\times Z_{2}$ symmetry occurs. 
Considering the phase diagram of the quantum Ashkin-Teller 
and the Haldane's conjecture, we predicted in our previous paper 
that for arbitrary $S$ XXZ spin chains with bond-alternation, 
there are $2S+1$ BKT lines, $2S$ 2D Gaussian lines, and $2S+1$ 2D Ising 
lines in the region $-1<\delta<1$(summarized in table 3).
For $S=3/2$ isotropic case $\Delta=1$, 
Yajima and Takahashi\cite{Yajima96} studied 
with the density matrix renormalization group method 
evaluate the transition point as $\delta=0,\pm0.42$.


\section*{Acknowledgments}
We thank Professor H. Shiba and Dr. K. Okamoto. 
K.N. also acknowledges Professor H. J. Schulz and Dr. M. Yamanaka 
for fruitful discussions. 
This work is partially 
supported by Grant-in-Aid for Scientific Research (C) No.08640479 
from the Ministry of Education, Science and Culture, Japan. 
The computation in this work has been done using the 
facilities of the Supercomputer Center, Institute for Solid 
State Physics, University of Tokyo.

\appendix

\section{Renormalization group equations}
We derive the renormalization group equations(\ref{kosterlitz}). 
Let us consider the following Euclidean action,
\begin{equation}
  {\cal{S}}= {\cal{S}}_{0} + \sum_{\alpha}\frac{\lambda_{\alpha}}{2\pi}
  \int\frac{d^{2}\mbox{\boldmath$r$}}{a^{2}}
  {\cal{O}}_{\alpha}(z,\bar{z}),
\end{equation}
where ${\cal{S}}_{0}$ is a fixed point action, $a$ is a ultraviolet 
cutoff, $\mbox{\boldmath$r$} = (v\tau,x)$, 
and $z=v\tau+ix$, $\bar{z}=v\tau-ix$,. . 
We set the scaling operator ${\cal{O}}_{\alpha}$ as the normalized one as
\begin{equation}
  \langle{\cal{O}}_{\alpha}(z_{1},\bar{z}_{1})
  {\cal{O}}_{\beta}(z_{2},\bar{z}_{2})
  \rangle_{0} = 
  \frac{\delta_{\alpha,\beta}}{\left(\frac{z_{1}-z_{2}}{a}\right)^{2h_{\alpha}}
    \left(\frac{\bar{z}_{1}-\bar{z}_{2}}{a}\right)^{2\bar{h}_{\alpha}}},
\end{equation}
where $z_{1} = v\tau+ix$, $\bar{z}_{1} = v\tau_{1}-ix_{1}$ and $h$ and $\bar{h}$ 
are the conformal weights and the scaling dimension of ${\cal{O}}_{\alpha}$ is 
$x_{\alpha}=h_{\alpha}+\bar{h}_{\alpha}$. 

According to Zamolodchikov\cite{Zamo}, and Ludwig and Cardy\cite{Ludwig}, 
we have the following one loop renormalization group equations 
for the scaling transformation 
$a\rightarrow a^{'}=e^{dl}a$
\begin{equation}
  \frac{d\lambda_{\alpha}}{dl} = 
  (2-x_{\alpha})\lambda_{\alpha}
  - \sum_{\beta, \gamma}\delta_{h^{\alpha}_{\beta\gamma},\bar{h}^{\alpha}_{\beta\gamma}}
  \frac{C_{\alpha\beta\gamma}}{2}\lambda_{\beta}\lambda_{\gamma},
\label{rgeq}
\end{equation}
where $h^{\alpha}_{\beta\gamma}=h_{\alpha}-h_{\beta}-h_{\gamma}$, and 
$C_{\alpha\beta\gamma}$ is the operator product expansion coefficient of 
${\cal{O}}_{\alpha}$, ${\cal{O}}_{\beta}$, and ${\cal{O}}_{\gamma}$. 

For the double sine-Gordon model(\ref{dsg}), 
we define
\[
  {\cal{S}}_{0}=\frac{1}{2\pi K}\int d^{2}\mbox{\boldmath$r$}
  \left[ 
  \left(\frac{\partial\phi}{v\partial\tau}\right)^{2}
  +\left(\frac{\partial\phi}{\partial x}\right)^{2}\right],
\]
\[
  {\cal{O}}_{0} = \frac{a}{K}\left[ 
  \left(\frac{\partial\phi}{v\partial\tau}\right)^{2}
  +\left(\frac{\partial\phi}{\partial x}\right)^{2} \right],
\]
\[
  {\cal{O}}_{1} = \sqrt{2}\cos\sqrt{2}\phi,
\]
\[
  {\cal{O}}_{2} = \sqrt{2}\cos\sqrt{8}\phi,
\]
where ${\cal{O}}_{0}$ is proportional to the Lagrangian density. 
Using 
\[
 \langle\phi(z,\bar{z})\phi(0,0)\rangle_{0}=-\frac{K}{2}\log
  \frac{|z|}{a},
\]
and Wick's theorem, 
we can easily derive the following operator product expansions
\begin{equation}
  {\cal{O}}_{1}(z,\bar{z}){\cal{O}}_{0}(0,0)
  =-\frac{K}{2}\frac{1}{|z/a|^{2}}{\cal{O}}_{1}(0,0)
  +\cdots,
\end{equation}
\begin{equation}
  {\cal{O}}_{2}(z,\bar{z}){\cal{O}}_{0}(0,0)
  =-2K\frac{1}{|z/a|^{2}}{\cal{O}}_{2}(0,0)
  +\cdots,
\end{equation}
\begin{eqnarray}
  {\cal{O}}_{1}(z,\bar{z}){\cal{O}}_{1}(0,0)
  &=& -\frac{K}{2}\frac{1}{|z/a|^{K/2-2}}{\cal{O}}_{0}(0,0) \\
  && +\frac{1}{\sqrt{2}}\frac{1}{|z/a|^{-K}}{\cal{O}}_{2}(0,0)+\cdots,
\nonumber
\end{eqnarray}
where `$\cdots$' dose not include ${\cal{O}}_{0}$, ${\cal{O}}_{1}$, 
and ${\cal{O}}_{2}$.
From these equations we have 
\[
  C_{011}= C_{101}=C_{110}=-\frac{K}{2},
\]
\[
  C_{022}= C_{202}=C_{220}=-2K,
\]
\[
  C_{112}=C_{121}=C_{211}=\frac{1}{\sqrt{2}}.
\]
Using these OPEs and eq.(\ref{rgeq}), 
we can derive the renormalization group equations(\ref{kosterlitz}).

\section{Correction of the Gaussian fixed line}
Our method in \S 3.2 to determine the Gaussian fixed line is affected 
by the descendant fields (critical dimension $x=K/2+2n$; $n$:integer) 
of the $\sqrt{2}\cos\sqrt{2}\phi$ field, since generally the self-duality 
(in the Ashkin-Teller language) is not exact for the finite size system. 
Considering correction terms from the descendant fields of $\sqrt{2}\cos\sqrt{2}\phi$ 
and $L_{2}\bar{L}_{2}${\bf 1}, 
$((L_{2})^{2}+(\bar{L}_{2})^{2})${\bf 1} ($x=4$) irrelevant fields, 
we obtain 
\begin{eqnarray}
  \lefteqn{x_{6,7}(L)} \\
  &&  = \frac{K}{8}\pm\frac{y_{1}}{2}\left(\frac{2\pi}{L}\right)^{K/2-2}
  \pm\sum_{n=1}^{\infty}\frac{c_{n}}{2}\left(\frac{2\pi}{L}\right)^{K/2+2n-2}
  +dL^{-2},
  \nonumber
\end{eqnarray}
since the OPE coefficient 
\[
\langle\sqrt{2}\cos\phi/\sqrt{2}| \sqrt{2}\cos\sqrt{2}\phi|\sqrt{2}\cos\phi/\sqrt{2}\rangle
\] 
changes sign under the transformation(\ref{duala}). 
Therefore, the crossing points behave
\begin{equation}
y_{1}^{\mbox{\small cross}}(L)
   = -\sum_{1}^{\infty}c_{n}
  \left(\frac{2\pi}{L}\right)^{-2n}.
\end{equation}

\clearpage

%
%

\begin{figure}
\epsfxsize=3.3in \epsfbox{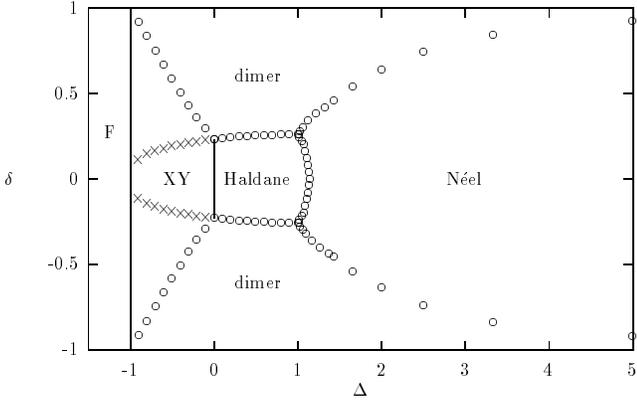}
\caption{Phase diagram in the $\Delta-\delta$ plane. 
The XY-dimer and the XY-Haldane phase boundaries are of the BKT type, 
the dimer-Haldane boundary is of the 2D Gaussian type, 
and the N\'eel phase boundaries are of the 2D Ising type. 
We also show the Gaussian fixed line ($y_{1}=0$) 
in the XY region ($\times$).
}
\end{figure}
%
\begin{figure}
\epsfxsize=3.3in \epsfbox{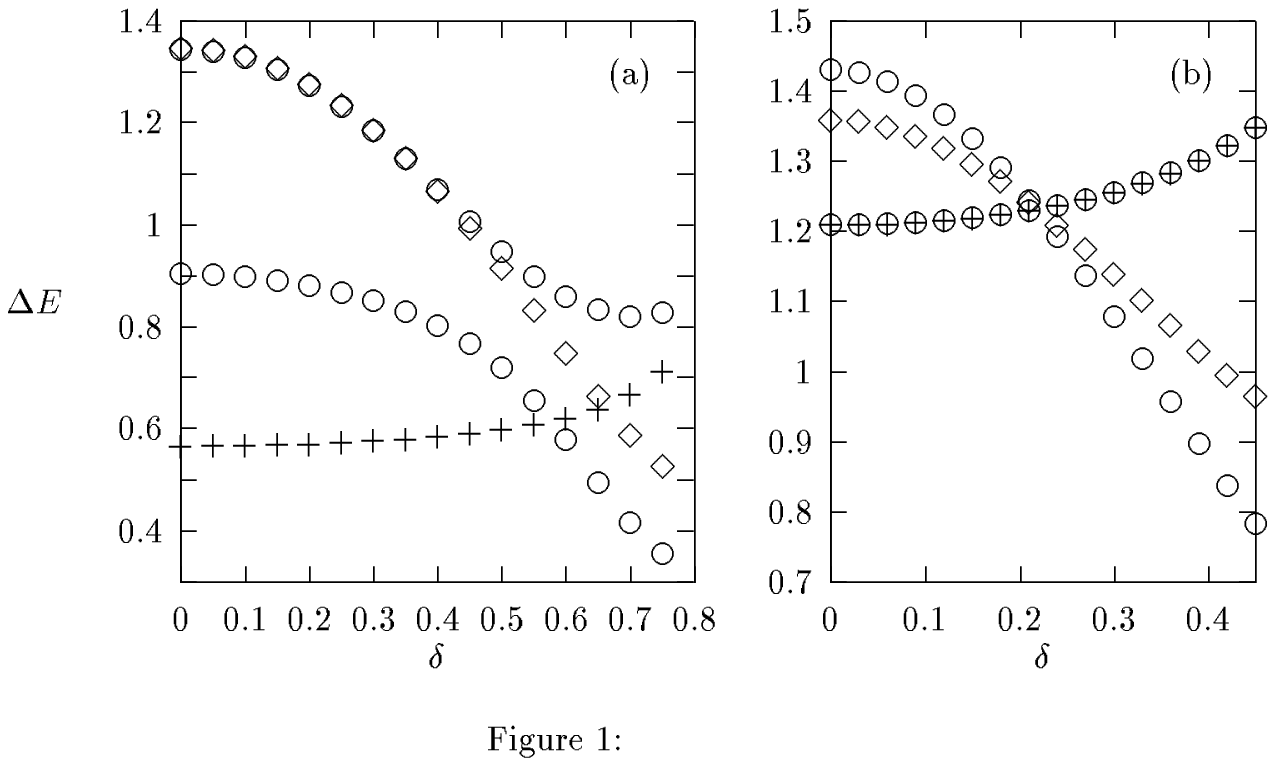}
\caption{Excitation energies for (a) $N=16$, $\Delta=-0.5$ 
and (b) $N=16$, $\Delta=0$. $\bigcirc$'s are $S^{z}_{T}=0$, $q=0$, $P=T=1$ 
excitations, $\Diamond$ is $S^{z}_{T}$, $q=0$, $P=T=-1$ excitations, 
and $+$ is the $S^{z}_{T}=\pm4$, $q=0$, $P=1$ excitation.
}
\end{figure}
%
\begin{figure}
\epsfxsize=2.8in \epsfbox{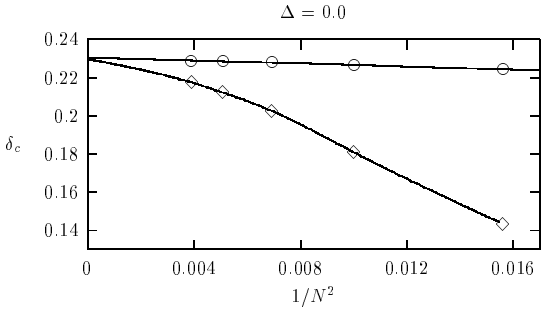}
\caption{The size dependence of the BKT multicritical point. 
$\Diamond$: the crossing point of ($S_{T}^{z}=0$, $k=0$, $P=T=-1$) and 
($S_{T}^{z}=\pm4$, $k=0$, $P=1$) excitations; Circles are obtained by 
the system with the twisted boundary condition(\S 3.2).
}
\end{figure}
%
\begin{figure}
\epsfxsize=3.3in \epsfbox{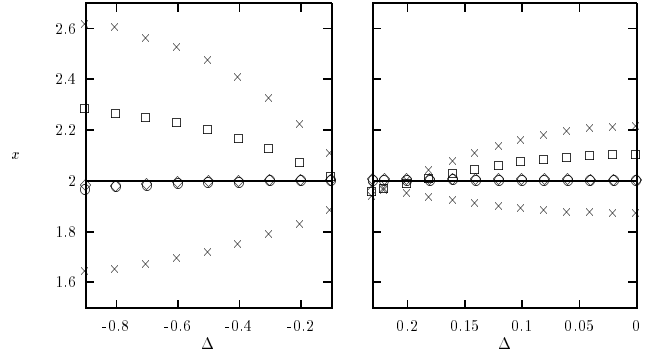}
\caption{The extrapolated values of $(2x_{0}+x_{1})/3$($\Diamond$) 
and $(x_{0}+x_{2})/2$($\bigcirc$) along the BKT transition lines. 
We also show the bare values of $x_{0}$($\times$), $x_{1}$($\times$), 
and $x_{2}$($\Box$) of the $N=16$ system.
}
\end{figure}
%
\begin{figure}
\epsfxsize=2.8in \epsfbox{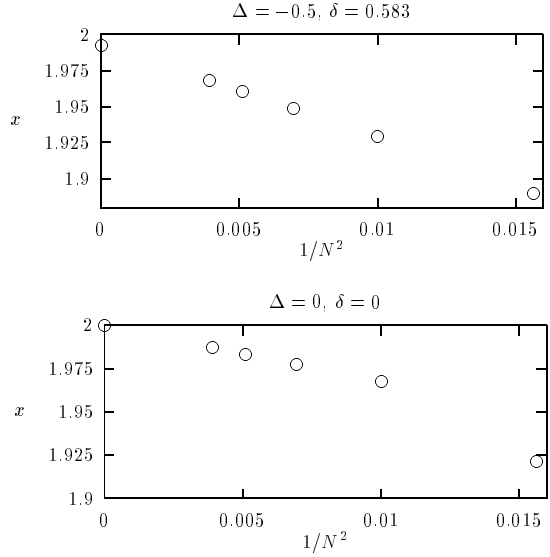}
\caption{The size dependence of ($3x_{0}+x_{1}+x_{2})/5$ as a function 
of $1/N^{2}$.This behavior comes from the $x=4$ irrelevant operators.
}
\end{figure}
%
\begin{figure}
\epsfxsize=3.3in \epsfbox{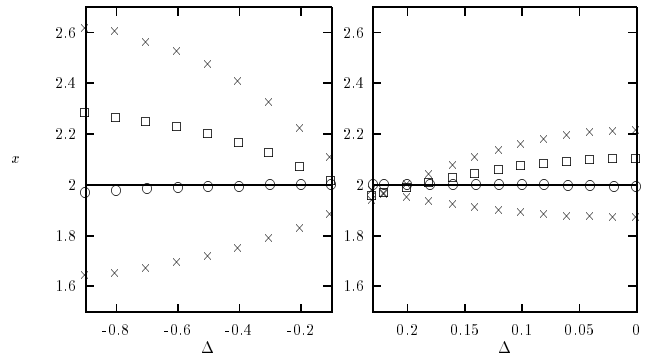}
\caption{The extrapolated values of ($3x_{0}+x_{1}+x_{2})/5$ ($\bigcirc$) 
along the BKT transition lines.
}
\end{figure}
%
\begin{figure}
\epsfxsize=3.0in \epsfbox{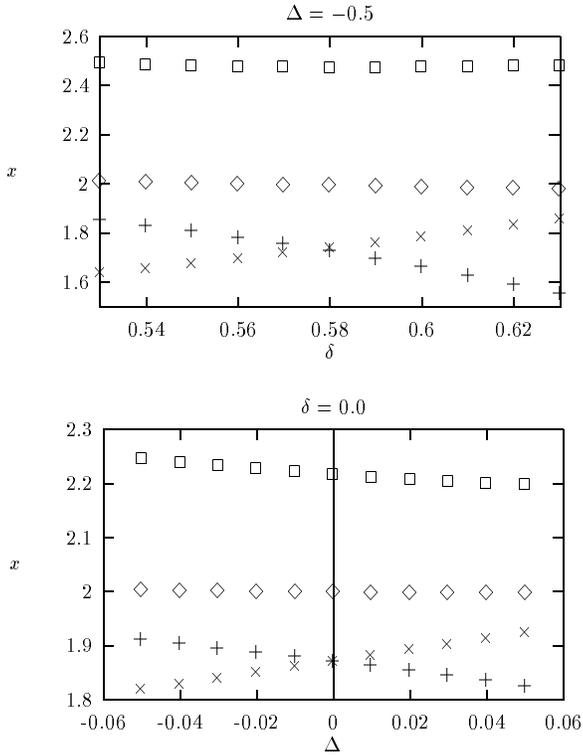}
\caption{The extrapolated values of $(x_{0}+x_{1}+x_{3})/3$ near 
the BKT critical point($\Diamond$). 
We also show the bare values of $x_{0}$($\times$), 
$x_{1}$($\Box$), and $x_{3}$($+$) of the $N=16$ system.
}
\end{figure}
%
\begin{figure}
\epsfxsize=2.4in \epsfbox{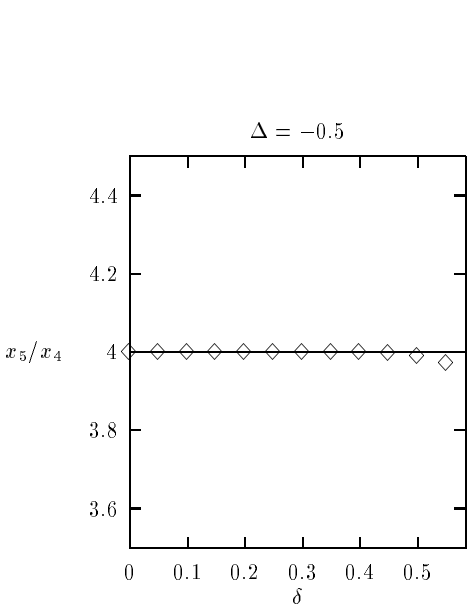}
\caption{The extrapolated values of $x_{5}/x_{4}$ in the XY phase. 
The transition point is $\Delta=-0.5$, $\delta=0.583$.
}
\end{figure}
%
\begin{figure}
\epsfxsize=2.4in \epsfbox{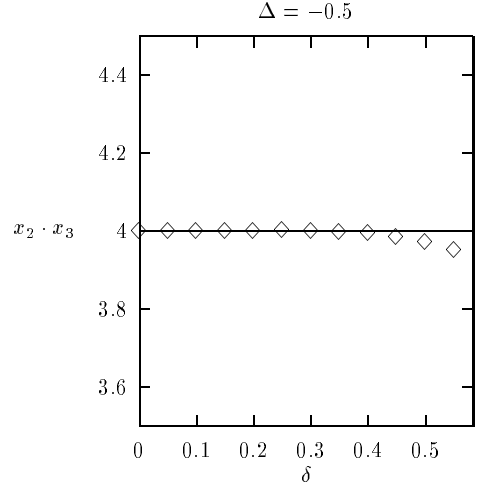}
\caption{The extrapolated values of $x_{2}\cdot x_{3}$ in the $XY$ phase.
}
\end{figure}
%
\begin{figure}
\epsfxsize=2.4in \epsfbox{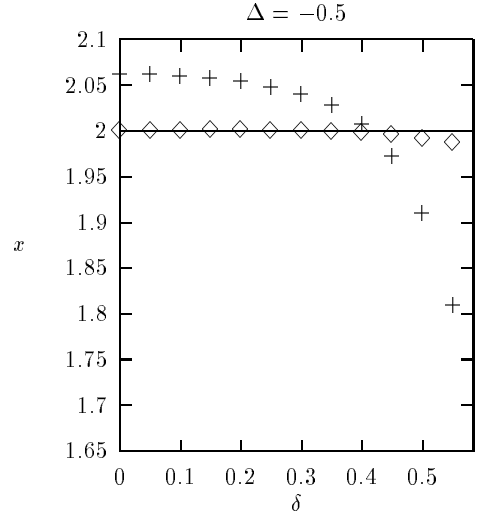}
\caption{$+$: The scaling dimension of the marginal operator of the 
$\Delta=-0.5$,$N=16$ system. 
$\Diamond$: The extrapolated values of $x_{0}+x_{1}-x_{2}$ 
which eliminates the logarithmic corrections. 
}
\end{figure}
%
\begin{figure}
\epsfxsize=3.3in \epsfbox{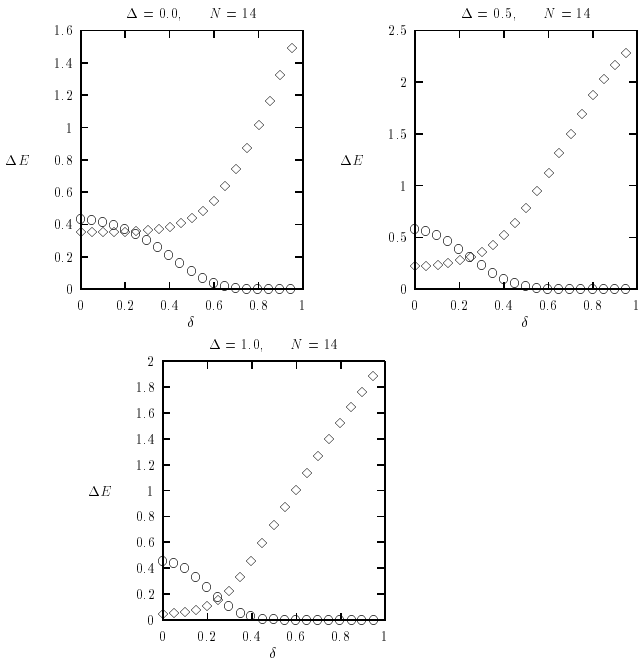}
\caption{Energy difference $E_{6}(\pi)-E_{g}(0)$ ($\bigcirc$) 
and $E_{7}(\pi)-E_{g}(0)$($\Diamond$). 
$\Diamond$ is the state with $\sum S^{z}=0$, $P=T=-1$, 
and $\bigcirc$ is the state with 
$\sum S^{z}=0$, $P=T=1$.
}
\end{figure}
%
\begin{figure}
\epsfxsize=2.4in \epsfbox{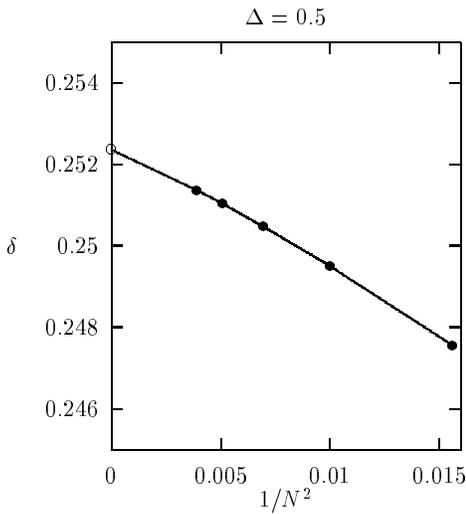}
\caption{Size dependence of the crossing points. 
The extrapolated value is $\delta_{c}=0.2524$.
}
\end{figure}
%
\begin{figure}
\epsfxsize=2.4in \epsfbox{ 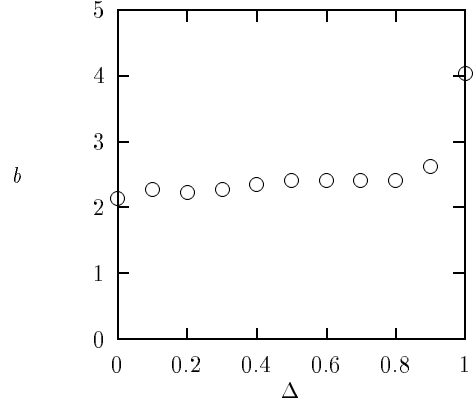}
\caption{The evaluated value of the exponent $b$ form eq.(25) 
for $N=14$.
}
\end{figure}
%
\begin{figure}
\epsfxsize=2.4in \epsfbox{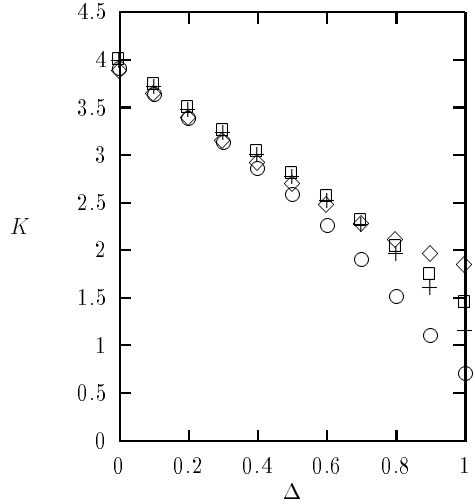}
\caption{The obtained values of $K$ for $N=16$ systems 
from $x_{1}$($\Diamond$), $x_{2}$($\bigcirc$), $x_{4}$($\Box$), 
and $x_{6}$($+$).
}
\end{figure}
%
\begin{figure}
\epsfxsize=2.4in \epsfbox{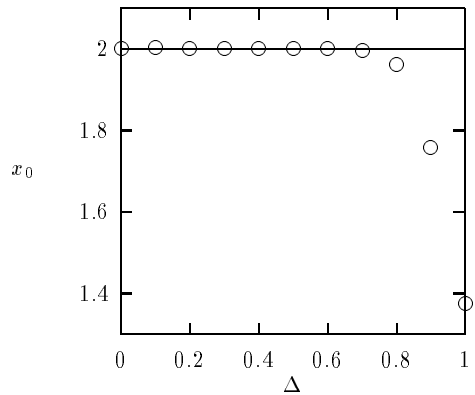}
\caption{The extrapolated values of $x_{0}$ which 
is the scaling dimension of the marginal operator.
}
\end{figure}
%
\begin{figure}
\epsfxsize=2.4in \epsfbox{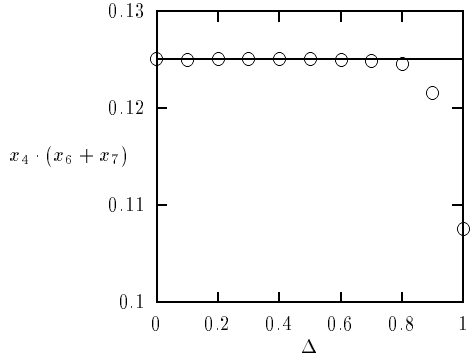}
\caption{The extrapolated values of $x_{4}\cdot (x_{6}+x_{7})$ on the 
Haldane-dimer critical line.
}
\end{figure}
%
\begin{figure}
\epsfxsize=2.4in \epsfbox{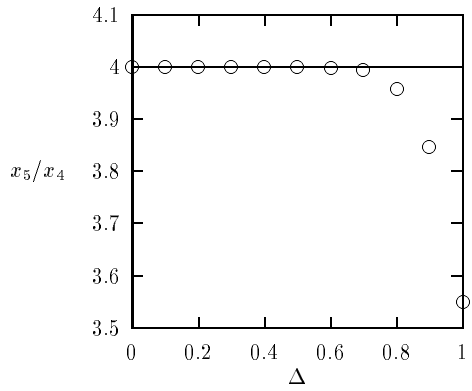}
\caption{The extrapolated ratio between $x_{5}$ and $x_{4}$.
}
\end{figure}
%
\begin{figure}
\epsfxsize=2.0in \epsfbox{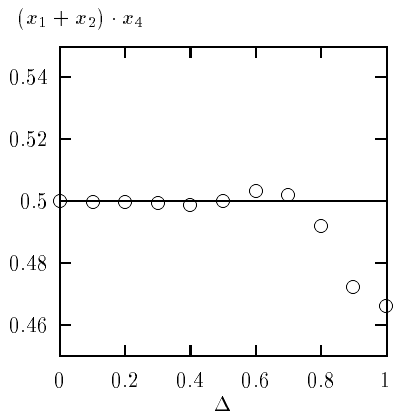}
\caption{The extrapolated values of $(x_{1}+x_{2})\cdot x_{4}$.
}
\end{figure}
\clearpage

\begin{figure}
\epsfxsize=2.3in \epsfbox{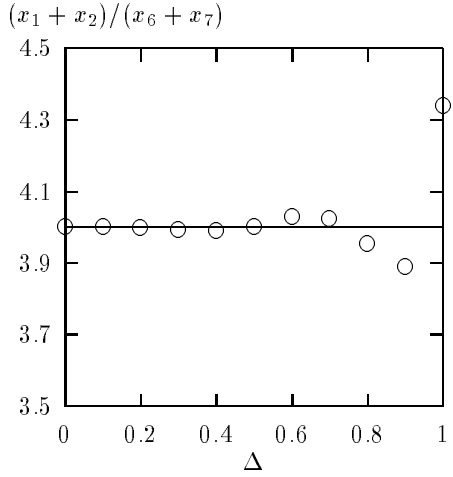}
\caption{The extrapolated values of $(x_{1}+x_{2})/(x_{6}+x_{7})$.
}
\end{figure}
%
\begin{figure}
\epsfxsize=2.8in \epsfbox{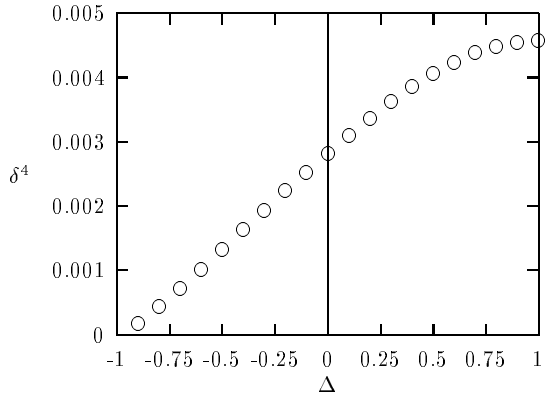}
\caption{The Gaussian fixed line in the $\delta^{4}-\Delta$ plane. 
$K$ takes a value from $\infty$ to $1$ in the region $\Delta:[-1,1]$. 
Near $\Delta=-1$, the Gaussian fixed line behaves 
as $\delta\propto(1+\Delta)^{1/4}$.
}
\end{figure}
%
\begin{figure}
\epsfxsize=2.8in \epsfbox{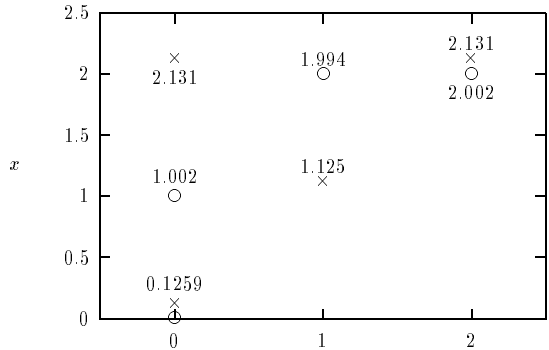}
\caption{The low lying scaling dimensions of the dimer-N\'eel transition point 
$\Delta=2.0$, $\delta=0.683$. 
Horizontal line is the momentum $\times L/2\pi$.  
$\bigcirc$ is the scaling dimensions of $P=T=1$ states, and 
$\times$ is of $P=T=-1$ states.
}
\end{figure}


\end{document}